\documentstyle[12pt]{article}
\def\bibref[#1]{\cite{#1}}
\def\p{{\scriptscriptstyle +}}
\def\m{{\scriptscriptstyle -}}
\def\be{\begin{equation}}
\def\ee{\end{equation}}
\def\C{C^{\alpha}_{\beta}}
\def\Z{Z^{\alpha}_{\beta}}

\def\vev#1{\langle #1 \rangle}

\def\darr#1{\raise1.5ex\hbox{$\leftrightarrow$}\mkern-16.5mu #1}

\def\half{{\textstyle{1\over2}}} 
\def\roughly#1{\raise.3ex\hbox{$#1$\kern-.75em\lower1ex\hbox{$\sim$}}}
\def\pl#1{{\it Phys. Lett.} {\bf #1B}}

\def\prd#1{{\it Phys. Rev.} {\bf D#1}}

\def\np#1{{\it Nucl. Phys.} {\bf B#1}}

\begin{document}
\thispagestyle{empty}
\begin{flushright} FERMILAB-PUB-94/137-T\\
NSF-ITP-94-50\\
hep-ph/9405374\\
\today
\end{flushright}

\vspace{6mm}

\begin{center}

{\Large \bf Fermion Masses from Superstring Models with Adjoint Scalars
\footnote{
Talk presented by J. Lykken
at the Second IFT Workshop on Yukawa Couplings and the Origins of Mass,
February 11-13, 1994, Gainesville, Florida.
}}\\ [6mm]
\vspace{25mm}

{\bf Shyamoli Chaudhuri}\footnote{
Work supported by the National Science Foundation Grants
PHY-91-16964 and PHY-89-04035.}
\\
{\it Dept. of Physics and Institute for Theoretical Physics\\
University of California, Santa Barbara, CA 93106}\\

\vspace{15mm}

{\bf Stephen-wei Chung and Joseph D. Lykken}\\
{\it Fermi National
Accelerator Laboratory\\
P.O. Box 500, Batavia, IL 60510}\\

\vspace{10mm}

\end {center}

\begin{abstract}
We explore the possibility of embedding
supersymmetric GUT texture ideas into superstring models.
We discuss the construction of GUT models using
free fermionic strings. We find $SO(10)$ models with
adjoint scalars, three generations of chiral fermions,
and a fundamental Higgs with a Yukawa coupling to
only the third generation.
\end{abstract}
\newpage

\section{Introduction}
Supersymmetric grand unified (GUT) texture models have had considerable
success reproducing the existing data for fermion masses and mixings,
starting from a restricted set
of effective operators at the GUT scale.
Two generic features of such schemes suggest that it may be
desirable, or even necessary, to eventually embed these structures
into full-fledged superstring models.

The first feature has to do with how SUSY GUT texture models generate
small numbers.
Fermion masses and mixings exhibit a number of distinct hierarchies,
characterized by ratios in the range roughly $1/10$ - $1/100$ or even
smaller, depending on how one (arbitrarily) chooses to parametrize them.
A popular idea in texture models\bibref[fn,dw] is that most of these
small numbers arise, not from small Yukawa couplings, but rather
from replacing Yukawas with higher dimension operators suppressed by
powers of
\be
{\vev{\Phi_{\rm adjoint}}\over M_X}\quad ,
\ee
where the numerator is the vev of an adjoint scalar, assumed to be
the GUT scale of about $10^{16}$ GeV, while $M_X$ is an even higher
mass scale, assumed to be roughly $10^{17}$ GeV. Thus, for example,
in one of the models of Anderson et al\bibref[ardhs], the following
$SO(10)$ invariant dimension $6$ operator resides in the $2$-$3$
component of the charged fermion mass matrices:
\be
O_{23} = 16_2\,10\,{45^2_{B-L}\over 45^2_1}\,16_3 \quad .
\ee
The superheavy scale $M_X$ is verging on the string scale, which is
estimated as
$M_{\rm string}$$\sim$$5 \times 10^{17}$GeV$\times g_{\rm string}$.
At any rate, the use of nonrenormalizable operators sensitive to
such high scales entails the risk of being overwhelmed by
``Planck slop''. Since superstring theory is the only known method
of controlling Planck slop, it may be necessary to invoke strings
in order to make valid statements about SUSY GUT textures.

The second feature of texture models which strongly suggests a
string interpretation are the textures themselves. The usefulness
of these schemes requires that the number of effective GUT scale
operators which make the dominant contributions to the fermion
mass matrices is rather few. In particular, the number of effective
operators up to, say, dimension $6$, should be considerably less
than the full set of effective operators allowed by the unbroken gauge
symmetries at the GUT scale. This requires that there are nontrivial
flavor-sensitive
selection rules, which provide the extra contraints and thus the
desired texture.
Superstrings are an obvious source for such selection rules.

In string models, couplings correspond
to correlators of vertex operators on
the worldsheet, and their vanishing depends on worldsheet symmetries.
Since many of these symmetries do not correspond to unbroken gauge
symmetries in spacetime,
the effective spacetime field theory below the string scale is
generically subject to a number of selection rules.
Of course, it is possible to avoid strings and obtain selection
rules from extra broken $U(1)$'s, discrete symmetries, $R$ symmetries,
and the like. However superstring theory not only provides similar
symmetries and selection rules, but also provides a more fundamental
understanding of them. This is true even though we at present have
no clue how nonperturbative string dynamics selects among the vast
number of perturbatively degenerate string vacua.

To see why, observe that
concrete phenomenological inputs can greatly narrow the range of
viable string vacua.
Given the SUSY GUT texture framework,
as well as improved low energy data,
one can obtain very specific guidance for string model building.
What is required is that one translate the phenomenological constraints
on the GUT scale effective Lagrangian into constraints on the world
sheet symmetries of the string models.
Then, for each particular
string model, which fixes a choice of the string vacuum, one can hope to
extract relationships between, say, the worldsheet structure that
guarantees three light generations of chiral fermions, and the
worldsheet structure that guarantees one of the observed hierarchies
in the fermion mass matrices. Any such relationships (if valid) would
be a profound new insight into particle physics. It is also
important to note that such relationships
depend only on order-of-magnitude computation, and thus do not
(necessarily) require the ability to make
precise determinations of string moduli.

\section{Superstring GUT's}
Historically, superstring model builders have made very little contact
with conventional GUT's. One of the reasons for this is that, in
string theory, gauge coupling unification occurs at the string scale
independently of whether or not matter fields assemble into
GUT multiplets\bibref[ginsparg].
Thus the agreement between LEP data and minimal
SUSY gauge coupling unification does not signify the existence of
GUT's in a string context\bibref[ib,fab,me].
Many quasi-realistic string models thus attempt to identify the
gauge coupling unification scale with the string scale, either by
raising the former or by effectively lowering the latter, and do
not obviously present two independent superheavy scales.

A technically important reason why so little effort has been put
into superstring GUT's is the fact that most string model
constructions simply do not allow the appearance of adjoint GUT scalars
in the massless spectrum at the string scale. Thus conventional
GUT symmetry-breaking vevs are not available. This is because in most
string constructions the GUT gauge group is realized at
Kac-Moody level one. The Kac-Moody level is a positive integer label
needed to specify unitary irreducible representations (irreps)
when Lie algebras are combined with worldsheet conformal symmetry to
produce Kac-Moody algebras. For fixed level there is a constraint
on the allowed irreps for massless matter fields.
Thus at level one the allowed irreps of $SU(5)$ and $SO(10)$ are
given by:
\begin{eqnarray}
{\rm SU(5)} &:&  1,5,\bar{5},10,\bar{10}  \nonumber \\
{\rm SO(10)} &:& 1,10,16,\bar{16} \quad .\nonumber
\end{eqnarray}
A more precise statement is that massless adjoint scalars are
incompatible with the presence of massless chiral fermions in
level one string models\bibref[dhl].

It is possible to build level one string models with unconventional
structures which are similar to GUT's. The flipped $SU(5)$ model\bibref[fli]
is the most developed example of this; there the breaking of $SU(5)$ is
accomplished by vevs of a Higgs $10$, rather than the adjoint. Since the
$SU(3)$$\times$$SU(2)$ singlet in the $10$ has nonzero hypercharge,
flipped $SU(5)$ requires an extra $U(1)$ and a nonstandard treatment
of hypercharge. While it is not a pure GUT, flipped $SU(5)$ is a useful
prototype for quasi-realistic models in the free fermionic superstring
construction. Indeed we have borrowed some of its worldsheet structure
in building the superstring GUT's described below.

Yet another reason why superstring GUT's have been neglected is
the fear of exotics. A conventional superstring GUT requires
a string construction with Kac-Moody level at least two. Higher levels
are required if scalars in other non-fundamental irreps are
desired. For example, to obtain a massless $126$ of $SO(10)$
would require\bibref[fer] a string model with $SO(10)$ at level $\ge$ five.
Higher levels introduce the possibility that modular invariance
of the string model will require the presence of other exotic scalars
whenever, say, the adjoint or the $126$ are present. Indeed the
simplest Kac-Moody modular invariants, the left-right symmetric
diagonal invariants, require that all {\it allowed} irreps do in
fact appear in the spectrum. Thus one might worry that
for $SO(10)$ at level two the $54$ would appear in addition to the $45$,
while at level five the $54$, $120$, $144$, etc would appear in
addition to the $45$ and $126$. We will show below that this
expectation is false for fermionic string models, and thus that
there is no problem with unwanted exotics.

There are two basic examples of superstring GUT constructions in the
literature. The first, due to Lewellen\bibref[dhl], is a
free fermionic string model
based on the minimal embedding of
$SO(10)$ at Kac-Moody level two, with
adjoint scalars and chiral fermions. The second, due to
Font, Ibanez, and Quevedo\bibref[fer], is an orbifold construction
that realizes the GUT group $G$ at level $n$ starting with
$n$ copies of $G$ at level one. We do not know of any particular
reason to prefer one of these constructions over the other.
However, we have chosen to base our exploration of superstring
GUT's on Lewellen's free fermionic string model.

We have not addressed the question of how a GUT scale of $10^{16}$
GeV gets generated in our models, separate from the string scale.
Understanding the hierarchy of scales in string theory
presumably requires an understanding of strong dynamics.

\section{Model Building}
Four dimensional closed free fermionic string models are heterotic
superstring vacua described by a worldsheet lagrangian for $64$
real (Majorana-Weyl) free fermions, together with the bosons that
embed the 4-d spacetime, and ghosts. This construction is described
in detail in refs \bibref[klt,abk,ab,klst]; we will, for the most part,
follow the notation and conventions of refs \bibref[klst,dhl].
Models are conveniently specified by their one-loop partition
functions; these involve a sum over spin structures:
\be
Z_{\rm fermion} = \sum_{\alpha,\beta}\;\C\,\Z \quad ,
\ee
where the $\C$'s are numerical coefficients, while $\alpha$ and $\beta$
are $64$-dimensional vectors labelling different choices of
boundary conditions for the fermions around the two independent
cycles of the worldsheet torus.
For each real fermion there are two possible choices of boundary
conditions around a given cycle: either periodic (Ramond) or
antiperiodic (Neveu-Schwarz). However for fixed $\alpha$ and $\beta$
the real fermions always pair up into either Majorana or
Weyl fermions; if a particular Weyl pairing occurs consistently
across {\it all} $\alpha$ and $\beta$, then this pair can be
regarded as a single {\it complex} fermion.
For such complex fermions more
general boundary conditions -any rational ``twists''-
are then allowed\bibref[klt,ab,ckt]. A useful notation denotes a
pair of Ramond fermions as a $-1/2$ twist, while a general
$m/n$ twist indicates the complex fermion boundary condition
\be
\Psi \rightarrow {\rm exp}\left[2\pi i\,{m\over n}\right]\;\Psi\quad .
\ee

It is convenient to regard the
partition function as a sum over physical ``sectors'' labelled by
the $\alpha$'s. The contribution of any sector $\alpha$ to the partition
function contains a generalized GSO projection operator.
Up to an overall constant, this is given by:
\be
\sum_{\beta}\;\C\;{\rm exp}\left[-2\pi i\beta \cdot \hat{N}(\alpha )
\right]\quad ,
\ee
where $\hat{N}(\alpha )$ is the fermion number operator defined in
the sector $\alpha$. There are subtleties in the proper
definition of $\hat{N}(\alpha )$ for real Ramond fermions; these
are discussed in ref \bibref[klst].

Thus building a fermionic string model
amounts to choosing
an appropriate set of $\alpha$'s, $\beta$'s, and $\C$'s,
then performing the GSO projections to find the physical spectrum.
These choices are greatly constrained by the requirement of
modular invariance of the one-loop partition function; in addition,
higher loop modular invariance imposes a factorization condition
on the $\C$'s. Together these requirements imply that the $\{\beta\}$
are the same set of vectors as the $\{\alpha\}$, and that, if two
sectors $\alpha_1$ and $\alpha_2$ appear in the partition function,
then the sector $\alpha_1 + \alpha_2$ must also appear. These facts
allow one to specify the full set of $\alpha$'s and $\beta$'s by
a list of ``basis vectors'', denoted $V_i$.

Of the $64$ real fermions, $20$ are right-moving and $44$ are
left-moving. The first pair of right-movers are spacetime fermions
(corresponding to the two transverse directions in 4-d), while
the other $18$ right-movers are ``internal''. The requirement
of a worldsheet supercurrent contructed out of the right-movers
and the spacetime bosons is a consistency constraint on model
building. As a result there is always a sector -denoted $V_1$-
that contains massless gravitinos, corresponding to an $N$$=$$4$
spacetime supersymmetry before the GSO projection. After the
projection one may have $N$$=$$4$, $N$$=$$2$, $N$$=$$1$, or no
spacetime SUSY at all. We will only consider models with
$N$$=$$1$ spacetime SUSY.

Fermionic string models always contain an ``untwisted sector'',
with $32$ Neveu-Schwarz Weyl fermions. The untwisted sector
contains the graviton, dilaton, and antisymmetric tensor field.
It generally also contains
gauge bosons, and massless scalars including gauge-singlet moduli.

The first step in building a fermionic string model for a
SUSY GUT is to find an embedding of the GUT root lattice
in the left-moving fermions.
As discussed in ref \bibref[dhl], this means identifying the
root vectors with vectors of ``fermionic charges'' for $n$
complex left-movers, where $n$ is $\ge$ the rank of the gauge group.
The fermionic charge of the Neveu-Schwarz vacuum is $0$; it
is $\pm 1$ for an excited Neveu-Schwarz fermion/antifermion.
For the Ramond vacuum the charge is $\pm 1/2$, corresponding to
the two degenerate vacuum states. Thus using complex Neveu-Schwarz and
Ramond fermions, root vectors have components taking only the values
$0$, $\pm 1/2$, and $\pm 1$. The length-squared of each root
vector is equal to $2/k$, where $k$ is the Kac-Moody level.

Consider a particular example. A
level two embedding of $SO(10)$ using, let's say, only
Neveu-Schwarz and Ramond fermions, requires that we find
$5$ simple roots, each with length-squared one, whose inner
products reproduce the Cartan matrix of $SO(10)$, and whose
components take only the values $0$, $\pm 1/2$, and $\pm 1$.
Such embeddings exist for any number of complex fermions
$\ge$ $6$  (although $SO(10)$ has rank five, there is
no solution with only $5$ complex fermions).

The minimal embedding, using $6$ complex
fermions, has simple roots given by\bibref[dhl]:
\begin{eqnarray}
&(0,0,0,1,0,0),\quad(\half,-\half,-\half,-\half,0,0),  \nonumber \\
&(0,0,1,0,0,0),\quad(0,\half,-\half,0,-\half,\half),  \nonumber \\
&(0,\half,-\half,0,\half,-\half).    \nonumber
\end{eqnarray}
Since there is an additional vector orthogonal to the space
spanned by these roots, these $6$ fermions actually embed
$SO(10)\times U(1)$.

The maximal nontrivial embedding, using $10$ complex fermions, has simple
roots given by:
\begin{eqnarray}
&(\half,\half,-\half,-\half,0,0,0,0,0,0), \nonumber \\
&(0,0,\half,\half,-\half,-\half,0,0,0,0), \nonumber \\
&(0,0,0,0,\half,\half,-\half,-\half,0,0), \nonumber \\
&(0,0,0,0,0,0,\half,\half,\half,\half), \nonumber \\
&(0,0,0,0,0,0,\half,\half,-\half,-\half), \nonumber
\end{eqnarray}

We adopt the convention that the first $2n$ real left-movers of
our string models will correspond to the $n$ complex fermions
that define an embedding. Having chosen a particular embedding,
the next challenge is to find a set of basis vectors, consistent
with modular invariance and the GSO projections, such that we
generate sectors with massless vector states whose fermionic charges
reproduce the root vectors of the embedding. Assuming that we can
in fact produce all the gauge bosons of some GUT
group, we are then guaranteed that all physical states will
assemble into irreps of the GUT group. Note that it is a simple matter
to translate from the Dynkin basis to the basis defined by the
simple roots written as fermionic charge vectors. Thus we
can determine which irreps physical states belong to
by simply mapping their weights back to the Dynkin basis.

\section{Three Generations}
We have begun an exploration of level two $SO(10)$ and
$SU(5)$ fermionic string GUT models which employ Lewellen's
minimal embedding of $SO(10)$. So far this search has been
limited to models whose fermions are Neveu-Schwarz, Ramond, or
complex with $\pm 1/4$ twists. Since the introduction of other
rational twists {\it loosens} the modular invariance constraints,
we expect that we have merely scratched the surface of possible
models with this embedding.

We find that the requirement of three light chiral generations
( three $16_L$'s in $SO(10)$ ) is very restrictive in our models,
much more so than, e.g., the requirement of adjoint scalars.
This is not surprising, since other fermionic string model
builders have encountered the same difficulty\bibref[fli,fa,alr].
{}From an exhaustive search, we find that there exist {\it no}
examples of three generation level two $SO(10)$ models,
with the minimal embedding, using {\it only} Neveu-Schwarz and
Ramond fermions. Thus three generations requires fermions
with other twists.

A useful starting point for constructing three generation models
is the $Z_2\times Z_2$ symmetric orbifold structure
described by Faraggi in ref \bibref[orb]. As pointed out
in \bibref[orb,nan], this structure is present in all known
level one fermionic string models that have three generations.
The orbifold structure consists
of two $Z_2$ twists  $\theta_1$, $\theta_2$, acting
symmetrically on left and right-moving $[SO(4)]^3$ lattices.
{}From now on it will be very convenient to write $0$, $1$ for
Neveu-Schwarz and Ramond fermions, respectively. In this
notation:
\begin{eqnarray}
\theta_1  &=& (1100)(1100)(0000)  \nonumber \\
\theta_2 &=& (0000)(0011)(1100) \quad . \nonumber
\end{eqnarray}
When this structure is realized in a full fermionic string model,
the partition function can obviously be decomposed into
a sum of four pieces with respect to this orbifold: an
untwisted sector, and the three twisted sectors $\theta_1$,
$\theta_2$, and $\theta_1\theta_2$. For level one models,
there is a straightforward way
-the ``NAHE'' set- of constructing basis vectors
such that one chiral generation resides in each of the three
twisted sectors.
It is quite possible that there are many unrelated ways of
obtaining three generation fermionic string GUT's, but
we have so far found it convenient to adapt structures
similar to the NAHE set in our GUT models.

This orbifold
trick for obtaining three generations
is not, strictly speaking, compatible with
realizing $SO(10)$ at Kac-Moody level two. The left-moving
$[SO(4)]^3$ lattice cannot, of course, overlap with the
$12$ left-mover slots reserved for the $SO(10)$ embedding.
In addition, for a sector containing a massless chiral  $16_L$ fermion
{\it all} of the left-moving structure is rather severely
constrained by the requirements of $SO(10)$ and modular invariance.
Furthermore, even if one realizes the orbifold structure in
three twisted sectors (which one might as well take to be
three basis vectors), the additional basis vectors which
produce $SO(10)$ gauge bosons will not respect the
left-right symmetry of the $Z_2\times Z_2$ orbifold.

In spite of these difficulties, we have found ways to
simultaneously realize both $SO(10)$ level two and obtain
three generations. This is demonstrated
in the next section with a specific model.

\section{Features of Superstring GUT's}
Since we have only a sampling of models, and have made particular
choices of embedding, level, gauge group, and string construction,
it would be imprudent to try to make any general statements
about the properties of superstring GUT's. Instead, we will be
content making a few observations about features of our models.

As an example, we list below the basis vectors of a
particular level two $SO(10)$ fermionic string model.
Here $0$, $1$ denote real Neveu-Schwarz or Ramond fermions, and
$\pm$ denotes a real fermion which pairs with another $\pm$ real
fermion to make a complex fermion with $\pm1/4$ twist.
The $20$ right-movers are separated from the $44$ left-movers
by a double vertical line.
A vertical line
separates out the $12$ left-movers that embed the $SO(10)$ weights.
The first two right-movers are the spacetime fermions.

$V_0$ is required in all fermionic string models by modular invariance.
The $V_1$ sector contains the gravitino, as already discussed.
Superpartners of states in some sector $\alpha$ will be found
in the sector $V_1 + \alpha$.
The $45$ massless gauge bosons of $SO(10)$ level two are contained
in the untwisted sector, $V_2$, $V_3$, $V_4$, $V_2$$+$$V_3$,
$V_2$$+$$V_4$, $V_3$$+$$V_4$, and $V_2$$+$$V_3$$+$$V_4$.

\goodbreak
\begin{eqnarray}
V_0&=(11111111111111111111\|
111111111111\vert 11111111111111111111111111111111)
\nonumber \\
V_1&=(11100100100100100100\|
000000000000\vert 00000000000000000000000000000000)
\nonumber \\
V_2&=(00000000000000000000\|
111111110000\vert 11111111000000000000000000000000)
\nonumber \\
V_3&=(00000000000000000000\|
000000000000\vert 00001111111100000000000000000000)
\nonumber \\
V_4&=(00000000000000000000\|
110000111111\vert 11001100110011000000000000000000)
\nonumber \\
V_5&=(11100100010010010010\|
111100001100\vert 10101010101010111000000000000000)
\nonumber \\
V_6&=(11010010100100001001\|
111100001100\vert 101001011010010000010000\p\p\m\m\p\p\p\p)
\nonumber \\
V_7&=(11001001001001100100\|
111100001100\vert 11110000111100000000110000000000)
\nonumber \\
V_8&=(00110110110110000000\|
000000000000\vert 01010101010101000001000000000000)
\nonumber \\
V_9&=(00\p\m0\p\m0\p\p1\p\p11\p\p1\p\p\|
00000000\p\p\p\p\vert 00001111000011001001\p\p\p\p\p\p\p\p0000)
\nonumber
\end{eqnarray}
\goodbreak

This type of model can always be transformed into a similar
level two $SU(5)$ model, either by adding a basis vector or altering
existing ones. This is because the $24$ roots of $SU(5)$
contained in $SO(10)$ appear precisely in the gauge boson sectors
listed above that do not contain $V_3$. It is tempting to suppose
that such $SU(5)$ models may retain the fermion mass texture of
their $SO(10)$ parent models virtually intact.

Three generations of $16_L$ fermions are contained in $V_5$, $V_6$,
$V_7$, plus $3\times 6$ additional sectors obtained by adding
$V_2$, $V_4$, $V_2$$+$$V_3$, $V_2$$+$$V_4$, $V_3$$+$$V_4$, or
$V_2$$+$$V_3$$+$$V_4$ to these (the massless states in $V_3$$+$$V_{5,6,7}$
are GSO projected out). To describe the $[SO(4)]^3$ lattice of the
$Z_2\times Z_2$ orbifold structure, let $\{r_1\ldots r_{20}\}$ denote
the right-movers, and $\{l_1\ldots l_{44}\}$
the left-movers. Then the right
and left-moving
$[SO(4)]^3$ lattices consists of the fermions
\begin{eqnarray}
\{ r_4,r_5,r_7,r_8,r_{10},r_{11},r_{13},r_{14},r_{16},r_{17},r_{19},
r_{20} \} \nonumber \\
\{ l_{14},l_{16},l_{17},l_{18},l_{19},l_{20},l_{22},l_{24},l_{25},
l_{26},l_{27},l_{32} \} \nonumber
\end{eqnarray}
If we were to truncate this model to include only $V_0$, $V_1$,
and $V_{5,6,7}$, then it would be
apparent that three sets of chiral fermions
reside in the three twisted sectors of a $Z_2\times Z_2$ orbifold.
However this structure {\it does not} guarantee three generations
in the full model.
In fact it is difficult to avoid generating extra chiral fermions in
other twisted sectors. This has proven so far to be the most severe
constraint on model building.

The observable gauge group of this model is $SO(10)\times [U(1)]^4$.
The hidden sector gauge group is $[SU(2)]^3\times [U(1)]^2$.
The hidden sector
gauge group, hidden sector matter content,
and the number of extra $U(1)$'s, are very model dependent.
However it appears that the rank of the hidden sector gauge group
is always $\le$ $5$.

This model contains a $45$ of adjoint scalars contained in sector
$V_8$ and the seven other sectors obtained by adding the gauge boson
sectors to $V_8$. Although this model has $246$ different sectors
that contain massless particles (before the GSO projections), there
are no additional $45$'s and no $54$'s. This seems to be a
robust feature: singlets, $10$'s, and $16$'s proliferate in these
models, but $45$'s and $54$'s appear once, twice, or not at all.
This is easily explained by writing the weights of various irreps
in our fermionic charge basis. One finds that the $45$'s and $54$'s
have weights which require Neveu-Schwarz excited fermions and antifermions.
These make large contributions to the mass formula, and it then
becomes difficult to satisfy all the constraints while keeping
these irreps massless.

The above argument should generalize
for (at least) most higher level fermionic string models.  Thus
we conclude that these models do not have a generic problem
with exotics -i.e., large massless irreps tend {\it not} to occur.
This fact may also imply that fermionic string models cannot embed
$SO(10)$ GUT models which, e.g., solve the doublet-triplet
splitting problem\bibref[babu] by introducing
several $45$'s and $54$'s.

The most interesting feature of the fermionic string GUT's which
we have looked at so far, is that they naturally possess one
of the key properties of most SUSY GUT texture schemes.
Unlike level one models, our models quite often have only
a single set of Higgs in the fundamental which is allowed
by fermionic charge conservation\bibref[klt,joel]
to have a Yukawa coupling to
chiral fermions.
One then observes that this Higgs generates {\it at most}
one Yukawa coupling, that
of the (by definition) third generation:
\be
\phi_{10}\Psi_{16}^{(3)}\Psi_{16}^{(3)}
\ee

For example, in the model above, there is a single $10$
with weights distributed among the
untwisted sector, $V_3$, $V_4$, $V_2$$+$$V_4$, $V_3$$+$$V_4$, and
$V_2$$+$$V_3$$+$$V_4$. All of these states have a Neveu-Schwarz
fermion/antifermion excitation in $r_9$/$r_{12}$. Conservation of
fermionic charge then tells us that the only allowed Yukawa
coupling to this $10$ is the diagonal coupling to the
fermion generation in $V_6$. In this particular model that is
not the end of the story, because the $16_L$ in $V_6$ also
carries an extra $U(1)$ charge not carried by the Higgs $10$,
and so the third generation Yukawa is also killed.
However we have found that the question of whether
the third generation Yukawa
survives or not is very model dependent.

It is very gratifying to see the key feature of most texture schemes
appear so naturally in fermionic string GUT's. Furthermore
we see that this feature is intimately related to the structure
of the chiral fermion sectors which was needed to produce
three generations!

Level one models with three generations have similar
properties, although they typically contain several Higgs, each
with only one allowed Yukawa. One must also be extremely cautious
about attaching too much significance to Yukawas. String models
generate operators which are higher dimension but contain scalars
that can get Planck scale vevs. Such terms are then unsuppressed
and contribute to the GUT scale effective action just like an
ordinary Yukawa. Clearly we need a detailed analysis of higher
dimension operators in fermionic string GUT models before we
can draw any definite conclusions. Further work is also needed
to understand fermion mixings and the sources of masses for
the first and second generations.

\end{document}